\documentclass{PoS}
\usepackage{epsfig}
\usepackage{cite}
\usepackage{rotate}
\usepackage{graphics}

\newcommand{\ee}{e^{+} e^{-}}
\newcommand{\pp}{p \overline{p}}
\newcommand{\Qpp}{Q_{\pp}}
\newcommand{\Qdpp}{Q_{\pp}^2}  
\newcommand{\be}{\begin{eqnarray}}
\newcommand{\en}{\end{eqnarray}}
\newcommand{\bc}{\begin{center}}
\newcommand{\ec}{\end{center}}
\title{BaBar results on $e^+e^-\rightarrow \pp$ by means of ISR}

\ShortTitle{BaBar results on $e^+e^-\rightarrow \pp$ by means of ISR}

\author{Rinaldo Baldini Ferroli\\
 $\emph {representing the BaBar Collaboration}$\\
        Centro Studi e Ricerche Enrico Fermi, Roma, Italy\\
        INFN, Laboratori Nazionali di Frascati, Italy\\
        E-mail: \email{baldini@centrofermi.it}}

\abstract{BaBar has measured with unprecedented accuracy the $\ee \rightarrow \pp$ cross section from 
the threshold up to $\Qdpp \sim 20\; GeV^2/c^4$, finding out an unexpected cross section, with plateaux and 
negative steps.
Evidence for a ratio $ |G_E/G_M| >1$ has also been found  as well as a sudden 
variation in $|G_M|$ just above the threshold.}

\FullConference{International Europhysics Conference on High Energy Physics\\
                 July 21st - 27th 2005\\
                 Lisboa, Portugal}

\begin{document}

\begin{figure}[h]
\bc
\epsfig{file=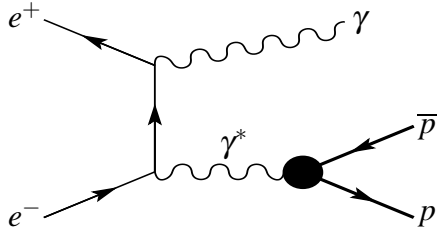,width=50mm,height=27mm}\vspace*{-3mm}
\put(0,0){\large$p$}
\put(0,33){\large$\overline{p}$}
\put(-25,73){\large$\gamma$}
\put(-75,26){\large$\gamma^*$}
\put(-155,-2){\large$e^-$}
\put(-155,73){\large$e^+$}
\caption{The diagram of the $\ee\rightarrow\pp\gamma$ process.}
\label{feyn}
\ec
\end{figure}

BaBar has measured with unprecedented accuracy the cross section $\sigma(\ee \rightarrow \pp \gamma)$ 
from the threshold up to a $\pp$ c.m. total energy squared $\Qdpp \sim 20\; Gev^2/c^4$ by means of the 
initial state radiation (ISR), as in Fig.\ref{feyn}. In fact it has been shown \cite{1} that at a $m_e /E_e$ 
precision level it is:
\be 
\frac{d\sigma}{d\Omega_\gamma dE_\gamma} (\ee \rightarrow \pp \gamma) = P(s,E_\gamma, \Omega_\gamma) 
\cdot \sigma(\Qdpp) , 
\label{cross-sectio}
\en
where $s$ is the $\ee$ total c.m. energy squared, $E_\gamma$ and $\Omega_\gamma$ are energy and angles 
of the ISR photon in the $\ee$ c.m. frame, $P$ is the density function for ISR emission, computed according 
to QED as in Fig.\ref{feyn}, and $\sigma(\Qdpp)$ is the  $\ee \rightarrow \pp$ cross section at the $\pp$ c.m. 
energy squared $\Qdpp$. 
%
%
\vspace*{-5mm}
\bc
\vspace*{-3mm}
\ec
%
%
In the Born approximation it is given by \cite{2}:
\be
\frac{d\sigma(\ee \rightarrow \pp)} {d\cos\theta} =  \frac{\pi \alpha^2 \beta C}{2 \Qdpp} 
\Big[(1+cos^2\theta)|G_M (\Qdpp)|^2 + \frac{4 M_p^2}{\Qdpp}sin^2\theta |G_E (\Qdpp)|^2\Big].
\label{cs-dphi}
\en 
All quantities are evaluated in the $\pp$ c.m. frame: $\theta$ is the azimuthal proton emission angle, 
$\beta$ is the proton velocity, $C$ is a  factor introduced in the final state distorted wave approximation 
to take into account the Coulomb interaction \cite{3}, $G_E$ and $|G_M|$ are the analytical time-like 
continuation 
of the corresponding electric and magnetic form factors (FF), as defined in the process 
$ e p \rightarrow e p $, 
where the exchanged $\Qdpp$ is space-like. Recently the knowledge of  the nucleon has been shaken to the 
roots by the new measurement of $G_E/G_M$ for space-like $\Qdpp$ \cite{4}, which is not constant at all, as 
believed for many decades, in spite of early theoretical predictions \cite{5,6}. At $\Qdpp = 4 M_p^2$ it is 
expected $G_E/G_M = 1$, assuming that  electric, magnetic, Dirac and Pauli FF are all analytical functions with 
respect to $\Qdpp$, hence continuous across the $\pp$ threshold. The same expectation is achieved assuming at 
threshold there is the $S$ wave only. Actually this expectation has been extended 
to the whole $\Qdpp$ explored range \cite{7,8,9,10,11,12,13},lacking high statistics 
measurements in particular concerning $G_E$. 
Actually what is quoted is $|G_M|$, also because at high $\Qdpp$ its contribution is the dominant one.
Concerning time-like $\Qdpp$ the first \cite{7}  and the most relevant measurements at low \cite{8} and at high 
$\Qdpp$ \cite{9}, previous to BaBar, has been performed by means of the inverse process $ \pp \rightarrow \ee $. 
Under the aforementioned hypothesis concerning $|G_M|$ they have shown a very steep increase approaching 
the threshold and a  $1/\Qpp^4$ overall behaviour, even earlier than asymptotically expected according to 
PQCD \cite{14}.

In the following ISR events have been selected by asking the ISR photon is detected, to get rid of the non ISR 
multihadronic background. A rather large fraction of events is lost in this way, however in this case BaBar has
the advantage over a conventional c.m. $\ee$ collider that the cross section can be measured even at threshold, 
with a $\sim 1\; MeV/c^2$ $\Qpp$ invariant mass resolution and  with almost full $\pp$ angular coverage. 
The B-factory PEP II ($9 GeV/c^2$ $e^-$ colliding with $3.1\; GeV/c^2$ $e^+$) and the BaBar detector have been described 
in detail several times \cite{15}. For the present purposes charged particle tracking (a 5 layer silicon vertex, 
SVT, and a 40 layer drift chamber, DCH) and identification ( by means of an internally reflecting ring imaging 
Cherenkov, DIRC, of ionization in SVT and DCH and energy deposition in the calorimeter) systems are the main 
components. Muons are identified by means of the instrumented iron flux return. The ISR photon is detected and 
must be in the range $22^o < \theta_\gamma < 137^o$ in the lab frame, however photon information has not been 
used in the kinematical fit.
The Monte Carlo (MC) event generator is based on the code described in \cite{16}. Extra ISR soft photons are 
generated according to the structure function method  \cite{17}.
%
%
\vspace*{-10mm}
\bc
\vspace*{-0mm}
\begin{figure}[h]
\bc
\begin{minipage}{155mm}
\begin{flushright}
\epsfig{file=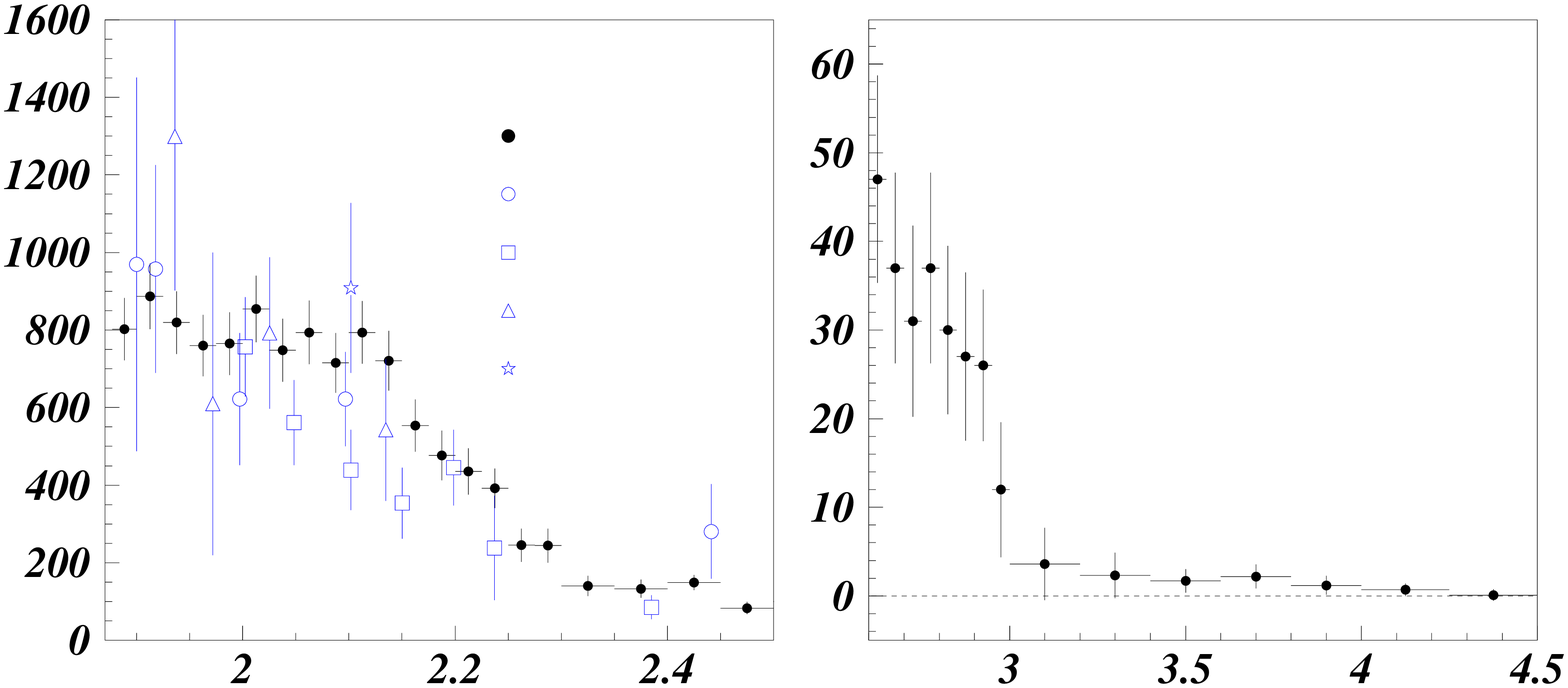,width=145mm}
\put(-235,170){\large (a)}
\put(-42,170){\large (b)}
\put(-428,98){\rotatebox{90}{$\sigma(\ee\rightarrow\pp)[pb]$}}
\put(-340,2){$\Qpp(GeV/c^2)$}
\put(-147,2){$\Qpp(GeV/c^2)$}
\put(-281,149){BaBar}
\put(-281,134){FENICE}
\put(-281,119){DM2}
\put(-281,104){DM1}
\put(-281,89){ADONE73}
\end{flushright}
\end{minipage}
\vspace*{-4mm}
\caption{The $\ee\rightarrow\pp$ cross section in comparison with previous experiments:
FENICE\cite{13}, DM2\cite{12}, DM1\cite{11} and ADONE73\cite{10}.}
\label{cs-data}
\ec
\end{figure}
\vspace*{-3mm}
\ec
%
%
At present a data sample related to  $232 fb^{-1}$ total integrated luminosity  has been analyzed.
Event selection has required an energetic photon and two opposite sign charged tracks, originated from the 
interaction point with a polar angle within the DIRC acceptance.
Radiative Bhabha events are  rejected if for each track it is $0.9 < E_{\rm cal} / P_{\rm tr} < 1.1$, between the 
energy deposition $E_{\rm cal}$ in the calorimeter and the DCH momentum $P_{\rm tr}$.
To get rid of the huge ISR $\pi^+ \pi^- \gamma$, $\mu \mu \gamma$, $K^+ K^- \gamma$ background both charged tracks 
are required to be well identified as proton candidates by means of a tight cut, leading to a loss of about 30\% 
of signal events.
A kinematical fit is performed adding further tight cuts, that is to be compatible only with proton masses 
hypothesis, leading to a further loss of about 25\%. The overall detection efficiency is about
18\% with a mild dependence on $\Qpp$.
In the surviving events there is no evidence of a peak at the $\rho$ mass or a peak at the $\Phi$ mass in a pion 
or kaon pair masses hypothesis and the estimated remaining contamination is negligible.
Angular and energy photon distributions are consistent with the ISR expectation. Final state radiation is 
expected to be very small and there is no interference term, due to the different charge parity. 
The most important and subtle source of background is the process $ \ee \rightarrow \pp \pi^0 $, where easily a 
soft photon is lost or the two photons are merged and not disentangled by the pattern recognition.
Since there are no experimental information on this process, events $ \ee \rightarrow \pp \pi^0 $ with a fully 
reconstructed $\pi^0$ are identified and the contamination in the selected sample of  $ \ee \rightarrow \pp$  
candidate events is evaluated according to the MC expectation. These estimated background events are about 5\% 
of the selected candidate events when $\Qdpp < 6\; GeV^2/c^4$, about 10\%  if $6 < \Qdpp < 9\; GeV^2/c^4$ 
and become consistent with 100\% above $\Qdpp \sim 20\; GeV^2/c^4$.

The ISR luminosity is calculated using the total integrated luminosity and the probability density function 
for ISR photon emission, as in eq.(\ref{cross-sectio}), taking into account the angular cuts. The luminosity, 
that has been integrated, depends on the $\Qpp$ invariant mass bin width, varying from $\sim 0.5 pb^{-1}$ at 
$\Qdpp \sim 2 \;GeV/c^2$ up to $\sim 1 pb^{-1}$ at $\Qpp \sim 3\; GeV/c^2$ for a $10\; MeV/c^2$ bin width.

Radiative corrections have been evaluated according the structure function method. They do not include corrections 
due to vacuum polarization. Hence what is quoted here is the so called ``dressed'' cross section.
The invariant mass resolution has been unfolded, however the chosen bin widths exceed the resolution.
With all these ingredients the calculated cross section $\sigma ( \ee \rightarrow \pp )$ is shown in Fig.\ref{cs-data}, 
statistical and systematic errors are quoted, including the uncertainties in detection efficiency, 
integrated luminosity and radiative corrections. For comparison some previous measurements are also shown.
The emerging cross section shape is an unexpected one: a flat plateau from the threshold up to
$\Qdpp \sim 5\; GeV^2/c^4$, followed by a step and then a second step at $\Qdpp \sim 8.5\; GeV^2/c^4$.
At present no simple explanation has been found concerning this behaviour.
\bc
\vspace*{-12mm}
\begin{figure}[h]
\begin{flushright}
\begin{minipage}{75mm}
\bc
\epsfig{file=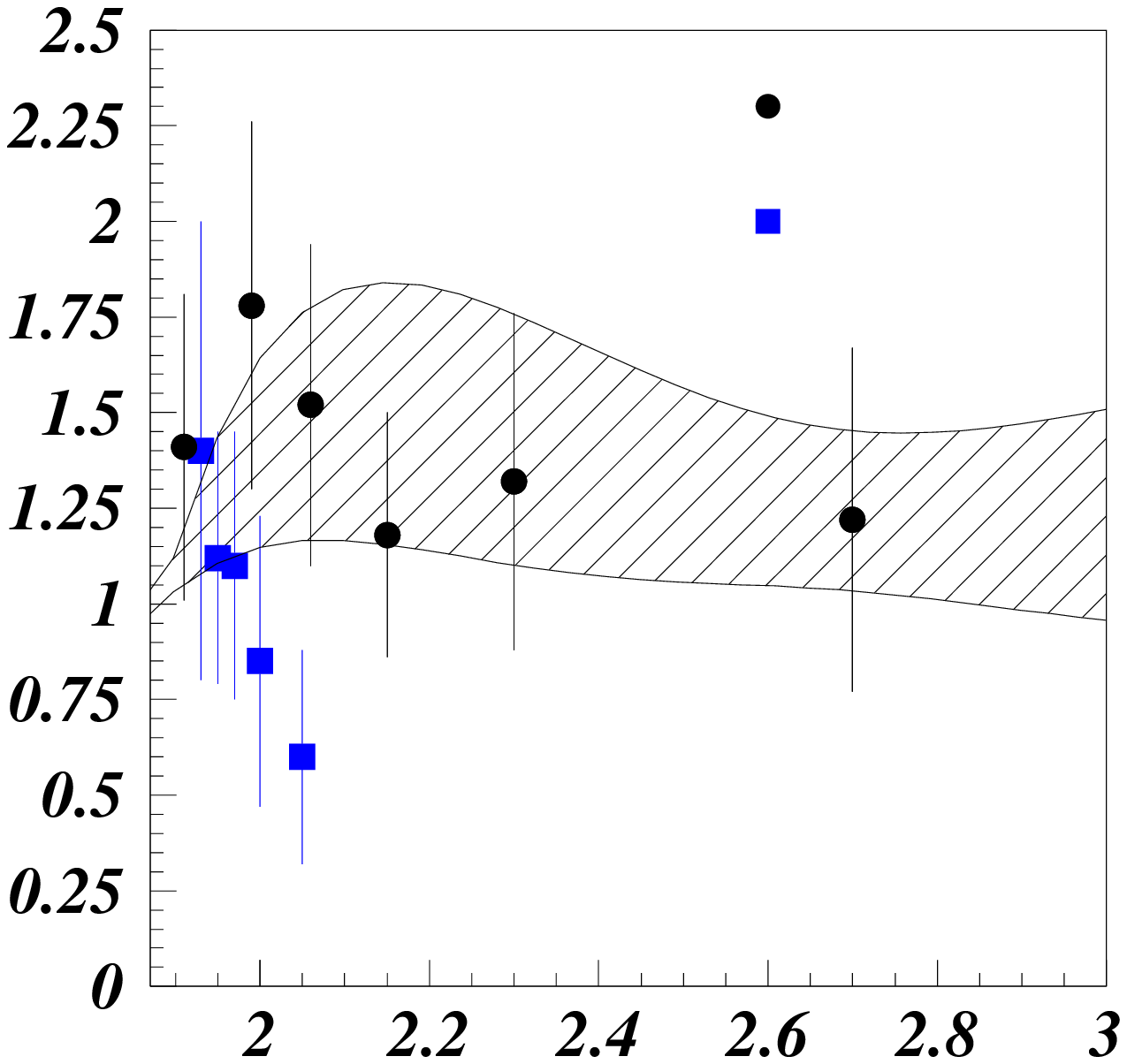,width=80mm}\vspace*{-1mm}
\put(-150,6){$\Qpp(GeV/c^2)$}
\put(-228,158){\rotatebox{90}{$|G_E/G_M|$}}
\put(-82,181){BaBar}
\put(-82,161){Lear}
\\
\begin{minipage}{65mm}\vspace*{-2mm}
\caption{The measured ratio $|G_E/G_M|$ in comparison with the Lear data\cite{8}.}
\label{ratio}
\end{minipage}
\ec
\end{minipage}
\hspace{-3mm}
\begin{minipage}{75mm}
\bc
\epsfig{file=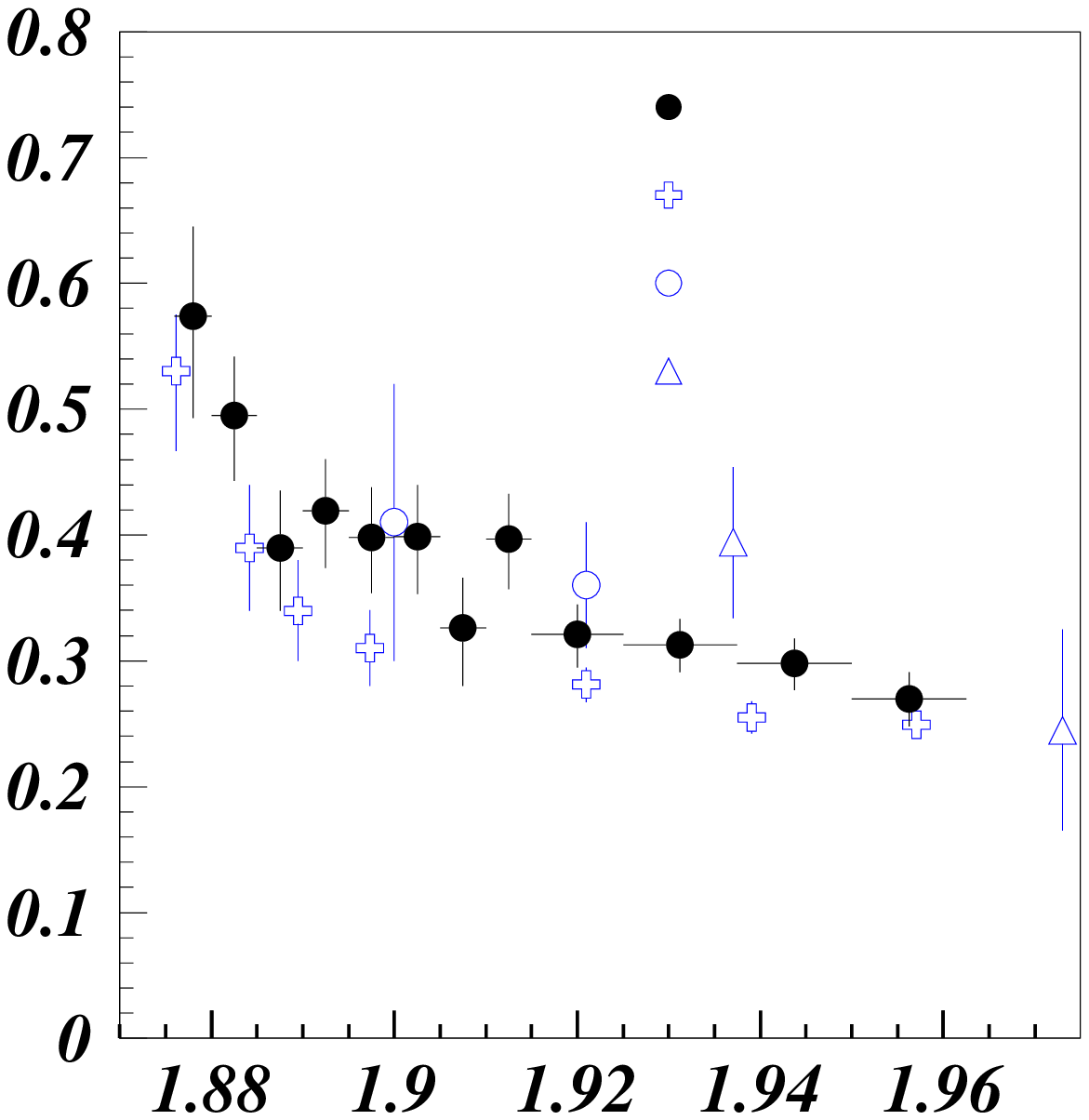,width=80mm}\vspace*{-1mm}
\put(-150,6){$\Qpp(GeV/c^2)$}
\put(-225,177){\rotatebox{90}{$|G_M|$}}
\put(-94,182){BaBar}
\put(-94,167){Lear}
\put(-94,152){FENICE}
\put(-94,137){DM1}
\\
\begin{minipage}{65mm}\vspace*{-2mm}
\caption{The proton magnetic form factor in comparison with other experiments.}
\label{gm}
\end{minipage}
\ec
\end{minipage}
\end{flushright}
\end{figure}
\vspace*{-5mm}
\ec
The angular distributions have been fitted, according to eq.(\ref{cs-dphi}) to extract the ratio $|G_E/G_M|$. 
The fitted ratios are reported in the Fig.\ref{ratio} . BaBar results are not in agreement with the APPLE \cite{8} 
results at LEAR, while they are in qualitative agreement with an expectation, updated, based on dispersion relations and on 
the JLAB space-like $|G_E/G_M|$ measurements \cite{18}. There is no clear evidence for an asymmetry, i.e. 
two photon exchange contribution, but there is not enough statistics to draw any conclusion.
As previous experiments did, $|G_M|$ is evaluated from the total cross section as shown in Fig.\ref{gm}, under the 
hypothesis $G_E = G_M$, which is at odd with the aforementioned results by the way. 
The asymptotic $ 1/(\Qdpp)^2$ behaviour found in $ \pp \rightarrow \ee $ is confirmed as well as a very steep 
slope very near threshold.
The $C$ factor in the cross section formula has a very steep slope too and is relevant only very near threshold,  
diverging as $1/\beta$ so that the cross section should be finite at threshold.
It has been introduced to get rid of the pointlike Coulomb interaction, since the FF definition should 
demand that the pointlike cross section has been factorized.
However this receipt concerning Coulomb interaction has been questioned \cite{19} and a better evaluation 
might affect the steep threshold behaviour of the FF.
It may be worthwhile to remind that no Coulomb corrections are expected in the $ \ee \rightarrow n \overline{n} $ process.
Alternative interpretations are peculiar $\pp$ final state interactions, like one $\pi$ exchange, as suggested 
in the $J/\psi$ radiative decay into a pseudoscalar $\pp$ pair \cite{20}, or a narrow $\pp$ vector state below 
threshold as obtained by means of dispersion relations in the unphysical region \cite{21}.

In conclusion BaBar has measured with unprecedented accuracy the $\ee \rightarrow \pp$ cross section from 
the threshold up to $\Qdpp \sim 20\; GeV^2/c^4$, finding out unexpected negative steps.
Evidence for a ratio $ |G_E/G_M| >1$ has also been found just above the threshold as well as a sudden 
variation in $|G_M|$.
\vspace*{-3mm}
\section*{Acknowledgments}\vspace*{-3mm}
I warmly acknowledge the chairman, S.Eidelman, and the BaBar Collaboration for 
having the opportunity\hspace{.55mm} to\hspace{.55mm} present\hspace{.55mm} these\hspace{.55mm} 
data\hspace{.55mm} and\hspace{.55mm} in\hspace{.55mm} particular \hspace{.55mm}
V.~P.~Druzhinin, S.~Pacetti, S.~I.~Seredniakov, E.~P.~Solodov and A.~Zallo for their 
inestimable and friendly collaboration.
%
%

\end{document}